\documentclass[sigconf,screen,nonacm]{acmart}

\usepackage{microtype}
\usepackage{balance}

\begin{document}
\begin{sloppy}
\title[Accountable Agents in Software Engineering: An Analysis of Terms of Service and a Research Roadmap]{Accountable Agents in Software Engineering:\\An Analysis of Terms of Service and a Research Roadmap}

\author{Christoph Treude}
\affiliation{%
  \institution{Singapore Management University}
  \city{Singapore}
  \country{Singapore}}

\renewcommand{\shortauthors}{Treude}

\begin{abstract}
AI coding assistants and autonomous agents are becoming integral to software development workflows, reshaping how code is produced, reviewed, and maintained. While recent research has focused mainly on the capabilities and impacts of productivity of these systems, much less attention has been paid to accountability: who is responsible when agents generate, modify, or recommend code? In practice, accountability is defined through the Terms of Service (ToS) and related policy documents that govern the use of AI-powered development tools.

In this vision paper, we present a comparative analysis of the Terms of Service for widely used AI coding assistants and agent-enabled development tools. We examine how these documents allocate ownership, responsibility, liability, and disclosure obligations between tool providers and software developers, and we identify common patterns and divergences between providers. Our analysis reveals a consistent tendency to shift responsibility for correctness, safety, and legal compliance onto users, as well as substantial variation in how providers address issues such as indemnification, data reuse, and acceptable use.

Based on these findings, we argue that existing policy frameworks are poorly aligned with increasingly agent-mediated and autonomous software development workflows. We outline a research roadmap for accountable agents in software engineering, identifying challenges and opportunities for modeling responsibility, designing governance artifacts, developing tooling that supports accountability, and conducting empirical studies of developers’ perceptions and practices.
\end{abstract}



\maketitle

\section{Introduction}
\label{sec:introduction}

AI coding assistants and autonomous agents are increasingly integrated into software development workflows, shaping how code is produced, reviewed, tested, and maintained.
Recent work has examined the capabilities of these systems and their impacts on developer productivity and code quality (e.g.,~\cite{peng2023impact,barke2023grounded,sergeyuk2025using,zhou2025exploring,dakhel2023github,bird2022taking}).
However, as assistants become more agentic (planning and executing multi-step changes, operating over larger code contexts, and acting with reduced direct supervision), questions of accountability become central: who is responsible when agent-generated changes introduce defects, vulnerabilities, or legal risks (e.g.,~\cite{asare2023github,yang2024swe})?

In practice, accountability is often articulated less through technical mechanisms than through governance artifacts such as Terms of Service (ToS), product terms, and policy addenda.
These documents specify ownership of inputs and outputs, disclaim warranties, allocate responsibilities, restrict acceptable use, and sometimes provide conditional indemnification.
Despite their practical importance, ToS are rarely treated as first-class objects of analysis in software engineering research.

In this paper, we analyze 14 policy documents (ToS and related documents) across nine AI coding assistants and agent-enabled development tools.
We compare how providers allocate (1) ownership and IP rights, (2) responsibility and liability for correctness, safety, and compliance, (3) data governance (e.g., training and retention), and (4) acceptable use constraints.

Our contributions are as follows: (1) A comparative analysis of how nine providers articulate ownership, responsibility, liability, and data governance in 14 policy documents; (2) a description of recurring policy themes (ownership, responsibility, indemnification, data reuse, acceptable use) and points of variation across providers; and (3) a research roadmap for accountable agents in software engineering.

\section{Background and Related Work}
\label{sec:background}

This section summarizes previous work on AI-assisted and agentic software development, with emphasis on empirical usage evidence, quality and risk considerations, and accountability and governance.

\subsection{AI assistants and agentic tooling in software engineering}
Empirical and qualitative research has started to describe how developers integrate code-generating assistants into everyday work.
Studies report that interaction is rarely a simple ``accept or reject'' decision; instead, developers iteratively steer suggestions, inspect generated code, and refactor output to fit local context~\cite{barke2023grounded, treude2025developers}.
At the team and organizational level, assistants have been associated with changes in task completion time and perceived productivity, but also with new needs for coordination and review~\cite{peng2023impact,bird2022taking}.
More recent work links these workflow changes to concrete development outcomes, including changes in the effects of efficiency, comprehension, and long-term maintenance~\cite{sergeyuk2025using,zhou2025exploring,pandey2024transforming,qiao2025comprehension,borg2025echoes}.

In parallel, research has shifted toward \emph{agentic} systems that plan and execute multi-step changes with less direct supervision~\cite{treude2025bot, he2025llm}.
These systems are typically evaluated on end-to-end software engineering tasks (e.g., issue resolution) rather than isolated completions~\cite{yang2024swe,zhang2024autocoderover}.
In addition to academic prototypes, practitioner-facing systems and announcements illustrate how quickly this space is moving~\cite{cognitionCognitionIntroducing}.

\subsection{Quality, security, and legal risk}
Although assistants can accelerate development, generated code can be incorrect, insecure, or otherwise unsuitable without careful review.
Work on security examines whether AI-assisted coding changes the prevalence and nature of insecure patterns and vulnerabilities, and under what conditions these failures occur~\cite{asare2023github,perry2023users}.
Recent studies discuss the broader risks and limitations of adopting AI-generated code in real projects, including quality trade-offs, maintenance concerns, and how practitioners manage these risks in situ~\cite{dakhel2023github,chong2024artificial}.

Legal and compliance risks are also substantial.
Concerns about training data, memorization, and reuse motivate questions about whether generated code may resemble copyrighted or licensed material, and how similarity across users should be interpreted~\cite{bender2021dangers,lee2023language}.
Empirical work on memorization and extraction provides grounding for these concerns and highlights mitigation strategies such as dataset deduplication~\cite{carlini2022quantifying,carlini2021extracting,kandpal2022deduplicating}.

\subsection{Accountability and governance}
As tools become more capable and agentic, accountability becomes a socio-technical problem that spans providers, developers, and organizations.
Recent work on AI accountability and governance highlights how responsibility can be distributed (or shifted) between stakeholders, how governance mechanisms shape practical oversight, and what it means to operationalize accountability in deployed systems~\cite{novelli2024accountability,schiff2020principles}.
Additional perspectives focus specifically on the governance of generative AI~\cite{taeihagh2025governance}, the role of records keeping and transparency mechanisms in oversight~\cite{chappidi2025accountability}, socio-technical aspects of agentic AI~\cite{donta2025socio}, and the fragmented global landscape of AI governance~\cite{schmitt2022mapping}.
In addition, documentation artifacts such as datasheets, model cards, and FactSheets aim to clarify intended use, limitations, and responsibilities, thereby supporting auditing and accountability~\cite{gebru2021datasheets,mitchell2019model,arnold2019factsheets}.

In this landscape, Terms of Service (ToS), product terms, and policy addenda operationalize accountability choices for AI-powered development tools by defining ownership, responsibilities, disclaimers, and constraints.

\begin{table*}[t]
  \centering
  \caption{Corpus of tools/providers and policy documents reviewed (links as of February~12,~2026).}
  \label{tab:corpus}
  \small
  \setlength{\tabcolsep}{3pt}
  \begin{tabular}{p{0.18\textwidth} p{0.32\textwidth} p{0.10\textwidth} p{0.34\textwidth}}
    \toprule
    \textbf{Provider / tool} & \textbf{Document} & \textbf{Date} & \textbf{URL}\\
    \midrule
    OpenAI (ChatGPT \& API) & Terms of Use & 2026-01-01 & \url{https://openai.com/policies/terms-of-use/}\\
    OpenAI (ChatGPT \& API) & Service Terms & 2026-01-09 & \url{https://openai.com/policies/service-terms/}\\
    \midrule
    GitHub Copilot & GitHub Terms for Additional Products and Features (Copilot section) & 2025-04-01 & \url{https://docs.github.com/en/site-policy/github-terms/github-terms-for-additional-products-and-features}\\
    GitHub Copilot & GitHub Copilot Product-Specific Terms & 2024-10 & \url{https://github.com/customer-terms/github-copilot-product-specific-terms}\\
    \midrule
    Anthropic (Claude) & Consumer Terms of Service & 2025-10-08 & \url{https://www.anthropic.com/legal/consumer-terms}\\
    Anthropic (Claude) & Acceptable Use Policy & 2025-09-15 & \url{https://www.anthropic.com/legal/aup}\\
    \midrule
    AWS CodeWhisperer & Service Level Agreement & 2023-04-04 & \url{https://aws.amazon.com/codewhisperer/sla/}\\
    \midrule
    JetBrains AI Assistant & JetBrains AI Service Terms & 2025-09-30 & \url{https://www.jetbrains.com/legal/docs/terms/jetbrains-ai-service/}\\
    JetBrains AI Assistant & JetBrains AI Terms & 2025-05-19 & \url{https://www.jetbrains.com/legal/docs/terms/jetbrains-ai/}\\
    \midrule
    Cursor (Anysphere) & Cursor Terms of Service & 2026-01-13 & \url{https://cursor.com/terms-of-service}\\
    \midrule
    Replit Ghostwriter / AI features & Replit Terms of Service & 2025-08-08 & \url{https://replit.com/terms-of-service}\\
    \midrule
    Sourcegraph Cody & Sourcegraph AI Terms & 2026-01-22 & \url{https://sourcegraph.com/terms/ai-terms}\\
    \midrule
    Google (Gemini Code Assist / Gemini API) & Google Terms of Service (Your content) & 2024-05-22 & \url{https://policies.google.com/terms}\\
    Google (Gemini Code Assist / Gemini API) & Gemini Code Assist Plugin Software License Agreement & 2025-11-05 & \url{https://developers.google.com/gemini-code-assist/resources/plugin-license}\\
    \bottomrule
  \end{tabular}
\normalsize
\end{table*}

\section{Corpus and method}
\label{sec:method}

This study treats Terms of Service (ToS) and closely related contractual documents as governance artifacts that operationalize accountability choices for AI-assisted development tools. Rather than analyzing marketing claims or technical documentation, we focus on binding policy texts that define ownership, responsibility, liability, data use, and constraints on use.

\subsection{Selection of tools and documents}
We derived an initial list of AI coding assistants from a design-space analysis of 90 systems across academia and industry~\cite{lau2025designspace}. To complement this source and ensure coverage of widely used industry tools, we also cross-referenced practitioner-oriented industry summaries of generative AI tools for software development. From these sources, we identified tools for which publicly accessible Terms of Service or comparable governance documents were available at the time of data collection. Some systems in the design-space study were no longer available, had been merged into other systems, or had been rebranded; the final corpus therefore consists of nine tools and fourteen governance documents. While not intended to be exhaustive, the corpus captures representative examples of AI coding assistants available at the time of data collection. For each provider, we include the primary contractual documents governing end-user use (e.g., Terms of Use/Service) and, where applicable, product-specific terms, AI-specific service terms, license agreements, and acceptable use policies. Our goal was not exhaustive legal coverage, but a consistent basis for cross-provider comparison of accountability-relevant clauses.

\subsection{Reviewed documents}
Table~\ref{tab:corpus} lists the documents reviewed, including the dates shown by the providers and the URLs last accessed on February~12,~2026. When multiple documents jointly govern an AI feature (e.g., general ToS plus product-specific AI terms), we treated the set of documents as a composite accountability framework for that provider. We report links and dates so that readers can verify the precise versions analyzed, since the policy text can change over time.

\subsection{Analysis approach}
Two researchers independently conducted a qualitative, comparative close reading of the corpus with an explicit focus on how accountability is allocated. Policy clauses were coded along four recurring dimensions: (1) ownership and intellectual property (inputs and outputs), (2) responsibility and liability (including warranties, disclaimers, accuracy warnings, liability caps, and indemnification), (3) data governance (retention, reuse, and service improvement), and (4) use and delegation framing (including language that anticipates automated or delegated interaction).

Coding was iterative: the researchers first identified recurring clause types (e.g., “you own outputs”, “use at your own risk”, “no warranties”, “indemnify and hold harmless”, “liability will not exceed”, “we may use content to improve the service”), then compared how each provider combined these clauses into a coherent allocation model. Ambiguous cases were discussed between researchers until agreement was reached; clauses that simultaneously disclaimed warranties and granted output rights were a recurring example. To illustrate, OpenAI's statement that users “own the Output” was coded under \textit{ownership/IP}, while the accompanying clause that users “must evaluate Output for accuracy and appropriateness” was coded under \textit{responsibility/liability}, capturing how the same provider simultaneously grants rights and allocates oversight obligations. Our intent is not to offer legal advice or doctrinal interpretation, but to analyze how contractual language structures socio-technical accountability in agent-mediated software development.

\section{Preliminary Findings}
\label{sec:findings}

In all reviewed documents, we observe a recurring pattern: providers grant users intellectual property rights in the content generated by the system (the “Output”), typically by assigning to users any rights, title and interest that the provider may have in the generated content. In practical terms, this means that users are contractually allowed to use, modify, distribute, and incorporate generated code or text into their own projects. At the same time, these rights grants are paired with clauses that place responsibility for the correctness, legality, and downstream consequences of using that output on the user. However, beyond this shared baseline, differences emerge in the way providers allocate risk, govern data, and frame liability.

\subsection{Output rights and user responsibility}
Several providers explicitly assign rights in output to users while simultaneously emphasizing that users bear responsibility for their use. For example, OpenAI states that users “own the Output” and that it “assign[s] to you all our right, title, and interest, if any, in and to Output”. At the same time, OpenAI warns that “use of our Services may, in some situations, result in Output that does not accurately reflect real people, places, or facts.” and that users “must evaluate Output for accuracy and appropriateness for your use case, including using human review as appropriate”. Ownership is thus paired directly with an obligation of independent verification.

Anthropic adopts a similar structure. It provides that “we assign to you all of our right, title, and interest---if any---in Outputs”, while prominently cautioning, in capitalized language, that “YOUR USE OF THE SERVICES, MATERIALS, AND ACTIONS IS SOLELY AT YOUR OWN RISK”. 

Replit likewise warns that “Code generated or suggested by our AI systems may be erroneous or incomplete” and that it “accept[s] no responsibility or liability for the accuracy of content on the Service”. JetBrains’ AI terms clarify ownership boundaries by stating that “You own the Inputs and Your Data”, while distinguishing these from “System-generated data [that] includes aggregate anonymized data on how JetBrains AI is used”. Ownership of content is preserved, but the responsibility for use remains with the submitting party.

Cursor (Anysphere) allocates responsibility primarily through indemnity language: users must defend and hold Anysphere harmless from claims arising from unauthorized use of the services, violations of applicable laws or third-party rights, or content they submit. The emphasis is less on output ownership and more on downstream accountability. Sourcegraph’s AI terms similarly clarify that “As between the parties, you own all Inputs to and Outputs generated by your use of Sourcegraph”, while situating use within broader limitation and liability provisions. 

Google’s Terms of Service state that users “retain ownership of any intellectual property rights that you hold in that content”, embedding AI functionality within a broader platform model in which rights retention coexists with standard service disclaimers. GitHub’s Copilot terms place full responsibility for AI suggestions on users, requiring them to ensure suggestions do not violate applicable law or infringe third-party intellectual property, privacy, or other rights.

\subsection{Risk posture}
Although disclaimers are widespread, providers differ in how directly they allocate downstream legal risk. For example, Replit requires users to “indemnify and hold Replit harmless from any loss or damage incurred by Replit as a result of your use of the platform”. Cursor similarly requires users to “defend and indemnify Anysphere” for claims arising out of unauthorized use, legal violations, or submitted content. Both adopt indemnification models that transfer considerable litigation exposure to the user.

JetBrains’ AI terms require users to “indemnify, defend, and hold JetBrains harmless” for claims relating to “Your Inputs and Outputs or the combination of Your Inputs and Outputs with other data”. Anthropic limits its exposure through liability caps stating that total aggregate liability “will not exceed \ldots the greater of the amount you paid \ldots and \$100”. This establishes a quantitative boundary on provider exposure.

OpenAI’s Service Terms provide output-related indemnification in enterprise contexts, yet exclude protection where customers “disabled, ignored, or did not use any relevant citation, filtering or safety features”. Protection is thus conditional on responsible use of safeguards.

\subsection{Data governance}
Another area of divergence is post-submission data use. For example, OpenAI states that it “may use Content to provide, maintain, develop, and improve our Services”, embedding improvement rights directly into the contractual framework. Replit connects public content to model development even more explicitly, stating that “Content published in public Apps may be used by Replit for improving the Service, including but not limited to developing or training large language models”. 

By contrast, JetBrains distinguishes between user-owned content and analytics data, stating that “System-generated data includes aggregate anonymized data on how JetBrains AI is used”, while reaffirming that “You own the Inputs and Your Data”. This creates a clearer conceptual boundary between content ownership and system telemetry. Anthropic similarly provides that “as between you and Anthropic \ldots you retain any right, title, and interest that you have in the Inputs”, while assigning Outputs to the user.

Google’s Terms clarify that users retain ownership of submitted content, while granting Google rights necessary to operate and improve services. The model is platform-centric rather than AI-specific, yet the structural pattern is similar.

\subsection{Error framing and bounded remedies}
Providers also vary in how directly they acknowledge AI fallibility and how tightly they constrain remedies. For example, OpenAI states that “use of our Services may, in some situations, result in Output that does not accurately reflect real people, places, or facts”. Anthropic provides that the Services and Outputs are delivered on an “AS IS” and “AS AVAILABLE” basis and that use is “SOLELY AT YOUR OWN RISK”. Replit warns that the Service “may contain errors, inaccuracies, or omissions” and disclaims responsibility for resulting loss. OpenAI’s Beta Services are offered “as-is” and are “excluded from any indemnification obligations”, reinforcing the experimental status of certain features.

\subsection{Automation awareness and delegated use}
Although most documents assume human oversight, their language is sufficiently broad to encompass automated or delegated interaction. For example, JetBrains provides that users are “solely responsible” for submissions and must ensure they possess all necessary rights to provide inputs. OpenAI prohibits users from representing that Output “was human-generated when it was not”, signaling explicit recognition of automated generation contexts.

Anthropic defines Inputs, Outputs, and “Actions” to include “software manipulation” and “system interactions” performed by the Services, thereby anticipating forms of delegated execution.

\subsection{Accessibility of governance signals}
Beyond the allocation of rights and responsibilities, the documents vary considerably in how accessible their governance signals are to software developers in practice. Most are legal texts written at a platform level, presupposing familiarity with contract law, liability caps, and indemnification structures. Key disclaimers are often embedded within lengthy general terms or emphasized through typographic conventions. All-capitalized warranty exclusions, for instance, are standard in law but unlikely to improve developer comprehension. The obligations these documents impose, particularly around output verification, licensing compliance, and indemnification, are rarely surfaced through product interfaces or developer documentation. This gap between contractual allocation and practical visibility is itself a research concern: even if accountability is formally specified, it may not be actionable for developers who encounter the governing terms only at account creation.

\subsection{Summary}
Across the nine providers, the documents consistently couple output-related rights with user responsibility for correctness and compliance, while diverging in indemnification breadth, data governance structure, and liability caps. The contractual architecture is human-centered: rights may be granted to users, but responsibility stays with them. This allocation is coherent for assistive use under close human supervision, but its adequacy becomes less clear as agents operate more autonomously, executing multi-step changes with reduced direct oversight. Whether the current contractual model remains appropriate when agents, rather than individual developer actions, initiate and execute code changes is an open question that the analyses in this paper motivate as a priority for future research.

\section{Research roadmap for accountable agents}
\label{sec:roadmap}

Accountability in AI-assisted software engineering is currently operationalized primarily through contractual risk allocation: providers grant output-related rights while emphasizing user responsibility, disclaiming warranties, and bounding exposure. This approach is coherent for assistive, human-supervised usage, but becomes increasingly strained as agents plan and execute larger changes with reduced direct supervision. Moving forward, research should focus on aligning accountability mechanisms with agentic reality by advancing responsibility modeling, governance-aware system design, accountability-supporting tooling, and empirical understanding of practice.

\subsection{Responsibility modeling in agentic workflows}
As our analysis shows, the ToS documents we examined treat agent behavior as an extension of user intent, effectively collapsing “delegation” into “use”: users own the output and bear responsibility for it, regardless of how much autonomous action the agent took to produce it. For agentic workflows, this is an incomplete abstraction: autonomy is gradual, and responsibility may need to be expressed across stages such as planning, proposing, executing, and verifying~\cite{ferino2026towards}. Software engineering research can contribute models that represent delegated authority and supervision boundaries in a way that is operationally meaningful. One concrete direction is to connect responsibility to specific artifacts and events in the workflow (e.g., distinguishing responsibility for initiating an action, approving a plan, executing a change, and merging or deploying it) so that accountability is not inferred only after a failure but is represented as part of normal development practice.

\subsection{Governance-aware and policy-aware agents}
The documents we analyzed constrain usage contractually, through acceptable use policies, indemnification clauses, and data governance provisions, but leave enforcement entirely to users and organizations with no technical mechanisms to operationalize these constraints. This gap between governance specification and technical execution creates an opportunity for software engineering research. As agents become more capable, there is growing opportunity to make governance constraints computable and actionable. Recent work shows that repository-level context files such as \texttt{AGENTS.md} and \texttt{CLAUDE.md} encode operational guidance, constraints, and project-specific policies in a configuration-like form~\cite{chatlatanagulchai2025agent, mohsenimofidi2025context, galster2026configuring}, and that \texttt{AGENTS.md} can improve agent efficiency without degrading task completion while being rapidly adopted across tens of thousands of repositories~\cite{lulla2026impact}. These findings suggest a path toward policy-aware agents that can interpret and operationalize governance signals embedded in both contracts and repository artifacts, while preserving meaningful human oversight in delegated workflows.

\subsection{Accountability-supporting tooling}
The contractual shift of responsibility onto users, which we observed consistently across all nine providers, assumes that users can validate outputs. Validation becomes harder as automation increases. Technical mechanisms can reduce this gap by making agent actions legible, reviewable, and auditable. In practice, this means tooling that records the provenance of agent-generated changes (including prompts, tool calls, and model/version identifiers), supports reliable audit trails that connect an eventual code state back to the sequence of agent actions, and differentiates review expectations based on origin and risk. These mechanisms do not replace contractual terms; rather, they make it realistic for developers and organizations to satisfy the responsibilities that ToS already impose, particularly when the volume and scope of agent activity exceed what traditional review habits were designed for.

\subsection{Input accountability and interaction dynamics}
The documents we analyzed treat inputs and outputs as distinct artifacts: users own their inputs, providers may use them for service improvement, and output rights are assigned to users. What the contractual framework does not address is how responsibility should be attributed across the full sequence of an agentic session: individual prompts, tool calls, intermediate reasoning steps, and final code changes. Research is needed on how input provenance should be captured and what obligations arise when inputs introduce bias, proprietary content, or security risks into the generation process. More broadly, the field should study how human-agent communication patterns evolve across extended sessions, and whether isolated prompt--response models adequately capture the accountability dynamics of sustained, context-dependent collaboration. What responsibilities LLM providers and users each bear for the fairness and safety of the full interaction chain---not just the final output---remains largely unaddressed by current governance documents.

\subsection{Empirical studies of practice and perception}
Finally, accountability is not only a policy and tooling question; it is also a question of how people interpret and operationalize responsibility in teams. Empirical research is needed to understand how developers read and internalize the ToS language, whether organizations adjust review and release processes when adopting agentic tools, and how responsibility is negotiated when failures occur. These studies can also examine second-order effects, such as whether explicit contractual disclaimers lead to more conservative tool usage, whether they create a false sense of security when output ownership is emphasized, or whether they push accountability work (testing, auditing, compliance) into roles that are not currently resourced for it.

\section{Conclusion}
\label{sec:conclusion}

The terms of service governing today's AI coding assistants place output rights and user responsibility on the same side of the contractual ledger: developers own what agents produce, and developers are accountable for verifying, licensing, and indemnifying it. This allocation made sense when the human was the primary actor and the model was a sophisticated autocomplete. As agents plan, execute, and deploy changes with growing autonomy, the contractual model lags the operational reality. The research agenda we outline, spanning responsibility modeling, governance-aware systems, accountability tooling, and empirical studies of practice, addresses this gap directly. The core challenge is not whether providers or users bear more legal risk, but whether current governance documents, written for assistive use, remain adequate as agents take on a larger share of the work. Addressing this challenge will require closer alignment between governance mechanisms and the realities of agent-mediated development workflows, so that accountability is not only contractually specified, but also operationally meaningful in practice.

\balance

\end{sloppy}
\end{document}